\documentclass[aps,prc,floats,floatfix,preprint,superscriptaddress,nofootinbib]{revtex4}

\usepackage{graphicx}
\def\be{\begin{equation}}
\def\ee{\end{equation}}

\begin{document}

\title{Model space truncation in shell-model fits}

\author{G.F. Bertsch}
\affiliation{Institute for Nuclear Theory and Dept. of Physics,
University of Washington, Seattle, Washington}
\author{C. W. Johnson}
\affiliation{Department of Physics, San Diego State University}

\date{shell.v7b.tex}

\begin{abstract}

We carry out an interacting shell-model study of binding energies 
and spectra in the $sd$-shell nuclei to examine the effect of truncation of 
the shell-model spaces.  Starting with a Hamiltonian defined in a larger space 
and truncating to the $sd$ shell, the binding
energies are strongly affected by the truncation, but the effect on 
the excitation energies is an order of magnitude smaller.
We then refit the matrix elements of the two-particle interaction to compensate for
the space truncation, and find that it is 
easy to capture 90\% of the binding energy shifts by refitting a
few parameters.   With the full parameter space of the two-particle
Hamiltonian, we find that both the binding energies and the excitation
energy can be fitted with remaining residual error about 5\% of the
average error from the truncation.  Numerically, the rms initial
error associated with our Hamiltonian is 3.4 MeV and the remaining
residual error is 0.16 MeV.  This is comparable to the empirical error found
in $sd$-shell interacting shell model fits to
experimental data\cite{br06}.

\end{abstract}

\maketitle

\section{Introduction}

The interacting shell model with fitted interactions is a 
powerful predictive tool of nuclear structure theory\cite{br88,ca05}, but there has
been little study of the error associated with truncation of the
shell model spaces.    The basic premise
is that there exists some Hamiltonian of the nucleon degrees of freedom
that accurately describes nuclear
structure properties.  The interacting shell model is then just a
numerical tool to calculate the properties of that Hamiltonian.  However,
except for the lightest nuclei, the spaces that one can include in
the calculation are much too small to obtain meaningful results using
the full Hamiltonian.  The prevailing philosophy for the
interacting shell model in heavier nuclei is to use an effective
Hamiltonian fitted to the set of nuclei under study.  Typically, 
the effective Hamiltonian includes one- and two-body matrix elements
augmented by an overall scaling with mass number $A$ to take into
account higher order operators in addition to the known systematic 
scaling of the rms radius. 

In this work we examine the role of space truncation with a model
study of spectra in $sd$-shell nuclei.  We first define a target Hamiltonian
that can be solved in a larger space, and take the eigenenergies as
the ``experimental" data to be fitted.  We then carry out the $sd$-shell
fits to the target spectra to examine the characteristics of the fitted
Hamiltonian.

 \section{Methodology}

A recent shell-model fit to $sd$-shell nuclei has been carried out
by Brown and Richter \cite{br06}.  The parameters in the effective
Hamiltonian are the 3 single-particle energies of the 
$(d_{5/2},d_{3/2},s_{1/2})$ orbitals
and their 63 interaction matrix elements.  In addition, there is a parameter
for the $A$-dependent scaling of the interaction matrix elements, which
however is not important for the fit\cite{brown}.
The experimental data that was fitted was the energies of 608 states
in 77 nuclei.  

For our model calculations, we shall try to follow ref. \cite{br06}
with respect to the size and character of the data set to be fit.
We consider the nuclei in the $sd$-shell for which
$Z \le N$, ranging from $^{17}$O to $^{40}$Ca.  This gives 90 binding
energies with respect to the closed $^{16}$O core.  Besides binding energies,
we also include the excitation excitations of the 6 lowest excited
states in the spectrum where they exist.  When there are fewer states
in the $sd$ space, 
eg. for the nuclei $(Z,N) = (8,9),(8,20)$, we take all the states of the
spectrum.  Altogether, we fit 559 excitation energies.

The target values of the energies, simulating the experimental data,
are calculated from a shell-model Hamiltonian in an extended
space.  The choice of the space requires a compromise between having
the Hamiltonian be computable with modest computer resources,
and having a space extension that will systematically affect the
properties across the range of nuclei under study.

We will then fit the energies of the large space calculation to
Hamiltonian parameters for the small space.  There a number of
aspects to the refit.  Most obviously, we ask how much the
rms residuals of the energy shifts shrink going from the original
parameters in the small space to a compete refit of all the
shell model parameters.  It is also interesting to see the extent that
simple parameters of the interaction can capture the main effects
of the shifts.  We make this quantitative by considering two
simple models of the interaction and asking how the rms residuals
shrink using only the parameters in the interaction model for the
refit.

\section{The target Hamiltonian}

For the target Hamiltonian, we begin with the full $sd$ Hamiltonian and
extend the space to include all $2-\hbar\omega$ excitations into
higher oscillator shells.  This permits 2-particle excitations
into the $pf$-shell and 1-particle excitations into the $sdg$-shell.  
This space is large but still reasonably calculable, with the
largest matrix dimension 50 million.

Within the $sd$ shell, the Hamiltonian matrix elements are given
by Brown and Richter's USDB (universal $sd$-shell interaction `B') \cite{br06}; 
we take the values for $A=18$ but do not rescale them for 
different $A$.  For the off-diagonal and the $pfsdg$ diagonal matrix elements,
we take a contact interaction with different strengths for the spin 0 and
1.  The strengths are chosen to roughly fit the largest matrix elements of
the USDB interaction.  The coefficients of the $\delta$-functions are
450 and 300 MeV-fm$^3$
for isospin zero and one, respectively; the
integrals  evaluated in harmonic oscillator wave functions with oscillator
parameter $\hbar \omega = 10.5$ MeV.  The single-particle energies are
taken to be 5 MeV for the $pf$ orbitals and 10 MeV for the $sdg$ 
orbitals, taken with respect to zero energy in the USDB Hamiltonian.
These energies are smaller than Hartree-Fock single-particle energies, but
the empirical spectroscopy of intruder states, namely the typical excitation energy of
odd-parity state,  requires a substantial
reduction of them. 

The quantities to be fitted are the differences in energies of the
USDB Hamiltonian and the extended-space Hamiltonian.
One technical difficulty is insuring that corresponding states
in the two Hamiltonians are properly paired.  Furthermore, in some nuclei
there are intruder states in the extended-space spectrum that lie
below the states that should be paired.
The intruders  are identified by examining the occupation probabilities of the
higher shells removed before making a correspondance with the $sd$-space
levels. The 
fraction of $pf$ and $sdg$ components is generally between a few percent 
up to about 20-30$ \%$ with our interaction.  The two states that exceeded 
this had probabilities
greater than 80\%, permitting an unambiguous identification as intruders.
They were excluded from the spectra to be fitted.

Carrying out this procedure, we find that the rms energy differences of the
full data sets is 3.4 MeV.  Considering binding energies and
excitation energies separately,  the rms differences are 9.2 MeV and
0.53 MeV, respectively.

\section{Least Squares Fit}

\begin{figure}  
\includegraphics [width = 11cm]{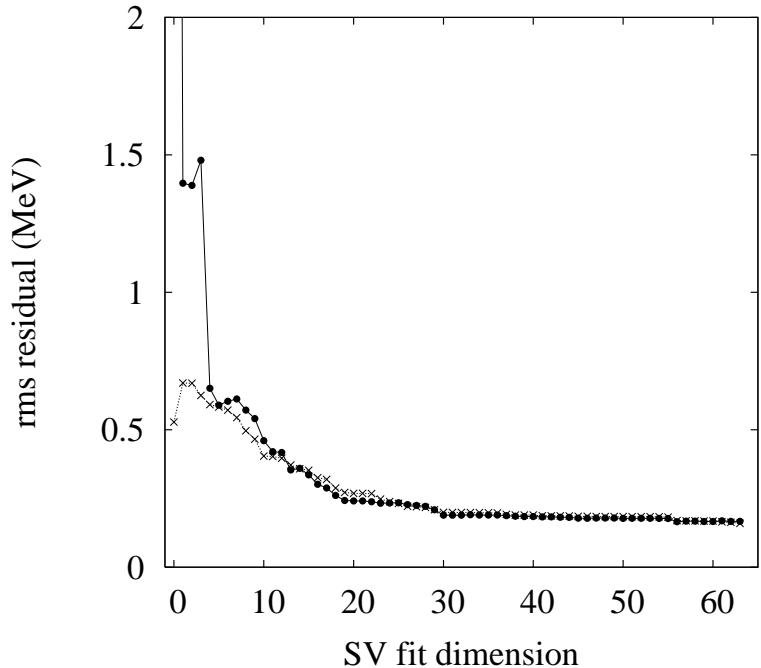}
\caption{\label{SVr} The rms residual error in the refitted energies
as a function of the parameter-space dimension in the singular
value decomposition.  Solid circles:  binding energies; crosses:
excitation energies.  The initial value of the binding energy residual 
is off-scale at 9.0 MeV.}
\end{figure}

\begin{figure}  
\includegraphics [width = 11cm]{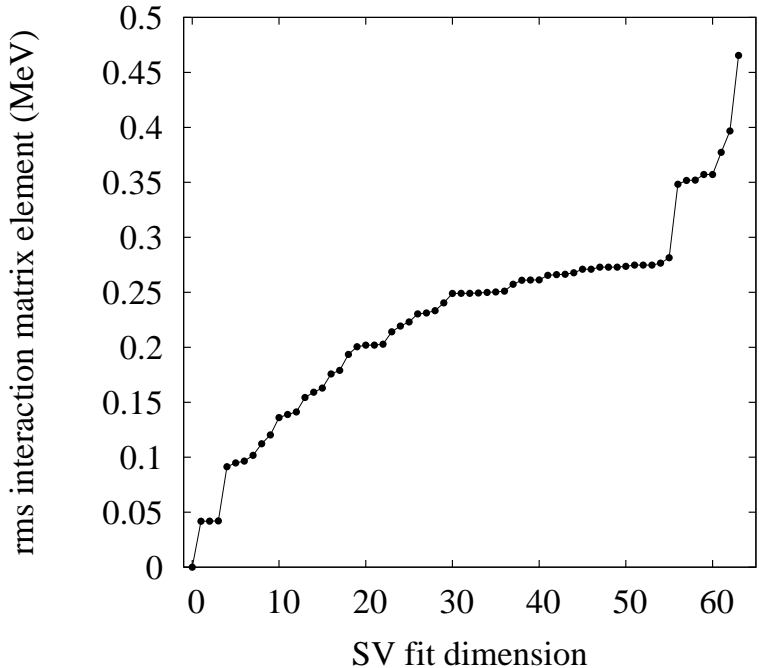}
\caption{\label{SVp} The rms change in the 63 two-particle interaction matrix element
as a function of parameter-space dimension in the singular value
decomposition.
}
\end{figure}

In data fitting, it is often the case that some parameters are
ill-determined and can assume large, unphysical values when one
carries out a linear least-square fitting procedure.  The standard remedy is
to make a singular value (SV) decomposition of the least-squares sensitivity
matrix, and monitor the quality of the fit as a function of the
rank of the SV decomposition.
The results are shown in Fig. \ref{SVr}, with rms averages for
binding energy and excitation energy residuals plotted separately.  We 
see that an enormous improvement in the binding energies can be achieved,
a factor of 40 of reduction of the rms error with 19 fitting parameters. By comparison, the improvement
in excitation energies is quite modest, reducing the rms error by a factor of 2 with
the rank 19 SV fit.  The quality of the fit becomes similar for binding
energies and excitations energies beyond rank 3.  The overall improvement
in the energies is a factor of 10 at rank 19 and 20 with high-rank fits.
However, the high-rank fits can be deceptive because the fitting procedure
relies on the Feynman-Hellman theorem to linearize the error matrix.
Qualitative information bearing on the linearization approximation may
be seen by plotting the rms changes in the interaction parameters as
function of SV rank.  This is shown in \ref{SVp}.
One sees that the interaction parameters change by an average of 0.2 MeV
for SV rank 19 and spike to 0.45 MeV at rank 63.  For a scale, the
rms average 2-body interaction in the original USDB parametrization is
1.9 MeV, suggesting that the interaction does not become ill-determined,
at least up to rank 30 or so.  We have confirmed the rank 30 results
by rediagonalizing the Hamiltonian matrix using the 
fitted interaction.  The resulting energies have an rms residual only
30\% higher than the SV value, confirming the utility of the linear
approximation up to that rank.


\section{Simple model Hamiltonians}

Since only very few parameters are needed to get the benefit of a
refit, the question naturally arises whether the key parameters can
be characterized in terms of simple Hamiltonians.  For example, it has 
been found that found that good results could be obtained
by refitting just the monopole part of a realistic interaction \cite{ca05,po81}; 
this phenomenology has been applied to a number of interactions \cite{ma97,ut99, ho04,su06}
(for details we point the reader to Eqns.~(9),(10) and Appendix B of \cite{ca05} and Eqn.~(4.1) 
of \cite{ma97}).
Another possibility, motivated by the Skyrme parametrizations \cite{sk56,be03}, and
effective field theory \cite{vk99}, is add an adjustable contact term to
interaction.  These two possibilities represent the extremes
of very long and very short for the characteristic ranges of the induced
interaction.  Let us see how well they work.

\subsection{Monopole interaction}
\begin{figure}  
\includegraphics [width = 11cm]{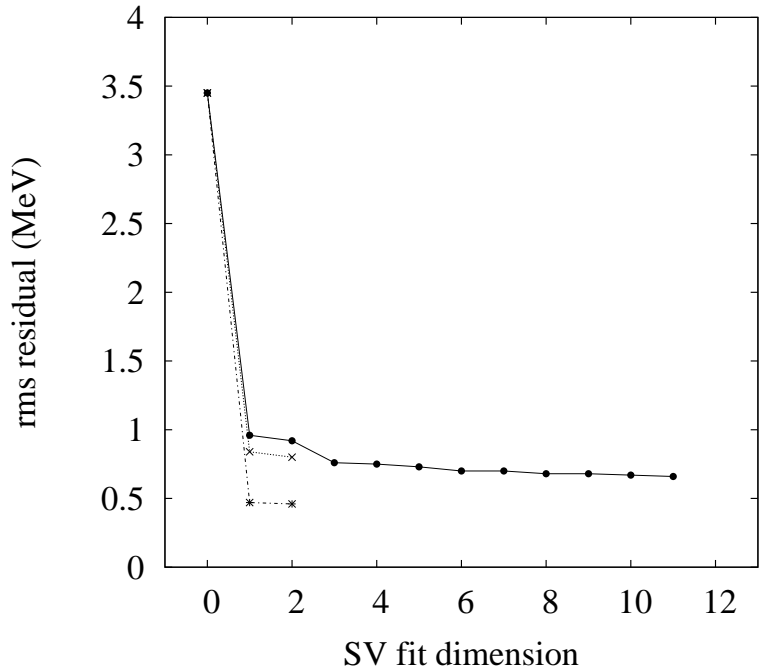}
\caption{\label{mono-delta} The rms residual error for simplified
forms of the refitted interaction.  Solid circles:  12-term
monopole interaction; crosses:  2-term monopole interaction;
stars: 2-term contact interaction.
}
\end{figure}

  The most general monopole interaction in the $sd$-shell has
separately adjusted coefficients for the 6 combinations of 
subshells as well as for the 2 isospins (T=0 and 1).  The
SV fits are shown in Fig. \ref{mono-delta} up to rank 11. 
One sees that the lowest rank captures a very large fraction
of the error, and higher ranks make incremental improvements.
Asymptotically, the remaining error is 0.7 MeV, which is still
4 times larger than what can be achieved in the 63-dimensional
two-particle matrix elements space.  To understand better the
nature of the refit producing the large initial error reduction,
we have also considered a monopole interaction independent of
subshells, i.e. an interaction depending only on particle number.
With two terms for the isospin dependence, we find the fits
shown by the crosses in the figure.  Interestingly, the 2-term
form gives better fits in the rank 1 and 2 spaces than the 
full monopole.

\subsection{Contact interactions}

There are two ordinary $\delta$-function
interactions, depending on spin ($S=0$ or 1).  The results of
the fit are shown as the stars in Fig.~\ref{mono-delta} and 
in Table I.  One
sees that it is much better than the monopole or the low-rank
SV vectors of the full $sd$ interaction space.

\begin{table}[h!]
\caption{RMS residuals (in MeV) in the SV decomposition with 
various treatments of the parameter fitting.
}
\begin{tabular}{|c|ccc|}
\colrule
& \multicolumn{3}{|c|}{No. of parameters} \\
Interaction & 2  & 11  & 63   \\
\colrule
$sd$ 2-body    & 0.80  & 0.40&  0.16 \\
full monopole & 0.92  & 1.18 & \\
reduced monopole & 0.84 &  &\\
contact & 0.46 & & \\
\colrule
\end{tabular}
\end{table}

\section{Conclusion}

With our model to study the effects of truncation on the spectra of
the interacting shell model, we obtained quite dramatic 
findings.  The energy shifts in the binding energies are more than
an order of magnitude larger than those in the excitation energies.
The binding energies shifts can be easily compensated by a 1-order
2-term interaction of a very simple form, either monopole or 
contact, with the contact interaction giving a better fit.
The importance of the monopole in the SV decomposition of
the sensitivity matrix was already shown in ref. \cite{pa06}.  In
that work, the authors considered the entire spectrum of fixed $J,T$
and $A$.  Our findings are similar, that a single operator close
to the monopole dominates the SV decomposition, but the ensemble
is very different--all the nuclei in the Fock space but only
a few levels in each nucleus.

After a few-parameter fit, the residual errors in
the binding energies and the excitation energies are comparable and
at the level of 0.5 MeV.   Further improvements can
be made in the full space of the 63 shell-model matrix elements,
with rank 30 achieving an rms residual of 0.2 MeV.  We note that
the limiting residual in the USDB  fit to experimental 
data\cite{br06} is  0.13 MeV.

However, it should be cautioned that these results may depend
on the specifics of the Hamiltonian model in the extended space.
The overall strength of the off-diagonal interactions cannot
be changed very much without either weakening the interaction
below what shell systematics require or enhancing it to an
extent where many intruder states would seriously contaminate the spectrum.
On the other hand, the spin dependence has been very much
oversimplified in our model and it could influence the fidelity
of the ultimate fit in the small space.  For example, it would be interest
to repeat the study including the tensor interaction,
which is known to have strong off-diagonal components.

In effective field theory \cite{vk99}, the machinery for performing 
truncations by the renormalization group is well-developed, with a systematic 
expansion ordered by counting powers of the relevant momenta, which 
allows one to estimate the error at a given order of 
truncation.  For the shell model Hamiltonian, there is no obvious analogy to 
power counting schemes, although one could argue that SV decomposition 
provides a logical framework for error estimates. Indeed we have seen that 
the linear approximation works very well for estimating the change in the
residual error. For example, as we go (somewhat
arbitrarily) from a rank-10 fit to a rank-30 fit using the SV decomposition,  
the change in the calculated energies has an rms value of 0.36 MeV, which is 
within a factor of 3 of the residual error in the complete fit. Besides 
giving an internal error estimate, such considerations might be helpful
in assessing the possible improvements of the fits by going to larger
spaces. Finally, it might be interesting to apply such analyses to other 
many-body approaches, such as energy-density functionals, where one can 
introduce a very large number of terms beyond the 10 or so present in the
most familiar parameterizations. 

It is intriguing that the error in the fit to experimental data, 0.13 MeV,
is actually smaller than the truncation error of our model Hamiltonian,
0.16-0.26 MeV.  It might be that the off-diagonal matrix elements in
the model Hamiltonian are too strong.  Certainly, the contact
interaction has larger off-diagonal matrix elements than a more
realistic interaction would have.  It also might be the case that
the extreme truncation in our model, only allowing $2\hbar\omega$ 
excitations in higher
shells,  requires stronger higher-order
operators than a more smooth truncation might produce.  In any case,
the closeness of the limiting error gives some hope that the accuracy
of the configuration-interaction theory might be improved by treating the 
higher shells.  

\section*{Acknowledgment}

This work was supported by the
UNEDF SciDAC Collaboration under DOE grants DE-FC02-07ER41457,  
DE-FG02-00ER41132, and DE-FC02-09ER41587.  We thank B.A. Brown, 
R. J. Furnstahl, and T. Papenbrock for discussions.

\end{document}